\documentclass[english,a4paper,eqno,11pt]{article}
\usepackage{amsmath}
\usepackage{amssymb}
\usepackage{cite}
\usepackage{latexsym}
\usepackage[english]{babel}

\font\tenbb=msbm10 at 12pt

\def\rR{\hbox{\tenbb R}}

\def\cal{\mathcal}

\font\titre cmbx10 at 18 pt

\def\gesp{\vskip1cm}
\def\esp{\vskip .6cm}
\def\pesp{\vskip .3cm}
\def\ppesp{\vskip 1mm}

\def\ni{\noindent}
\def\di{\displaystyle}

\def\Box{$\sqcap\hskip-.262truecm\sqcup$}
\newtheorem{thm}{Theorem}[section]
\newtheorem{defn}{Definition}[section]
\newtheorem{lem}{Lemma}[section]
\newtheorem{prop}{Proposition}[section]

\newtheorem{cor}{Corollary}[section]

\newtheorem{rem}{Remark}[section]

\usepackage{graphicx,epsfig,epic} 

\psfigdriver{dvips}

\usepackage{graphicx}
\usepackage{xypic}
\xyoption{all}

\begin{document}

\centerline{\titre{Deterministic Elaboration of Heisenberg's}}
\ppesp
\centerline{\titre{Uncertainty Relation and the Nowhere}}
\ppesp
\centerline{\titre{Differentiability}}
\esp
\centerline{\small Fay\c cal BEN ADDA$^1$, H\'{e}l\`{e}ne PORCHON$^2$}
\pesp
\centerline{\small \it $^1$New York Institute of Technology}
\centerline{\small\it  Campus 851, Road 3828, Block 388, Adliya, P.O. Box 11287, Bahrain}
\centerline{\small\it  benadda@ann.jussieu.fr; fbenadda@nyit.edu}
\centerline{\small \it $^2$University of Pierre and Marie Curie}
\centerline{\small\it  UFR 929, 4 Place Jussieu, 75005 Paris, France}
\centerline{\small\it  helene.porchon@etu.upmc.fr}

\gesp

{\small Abstract. In this paper the uncertainty principle is found via characteristics of continuous and nowhere differentiable functions. We prove that any physical system that has a continuous and nowhere differentiable position function is subject to an uncertainty in the simultaneous determination of values of its physical properties. The uncertainty in the simultaneous knowledge of the position deviation and the average rate of change of this deviation is found to be governed by a relation equivalent to the one discovered by Heisenberg in 1925. Conversely, we prove that any physical system with a continuous position function that is subject to an uncertainty relation must have a nowhere differentiable position function, which makes the set of continuous and nowhere differentiable functions a candidate for the quantum world.}\pesp


\esp

\esp

\section{Introduction}

The first formulation of the uncertainty relation was published by W.Heisenberg in 1927, \cite{HEI}, and expressed by the mathematical relation
\begin{equation}
\delta p\ \delta q\sim h
\end{equation}
where the quantities $\delta p$ and $\delta q$ were not explicitly defined (Heisenberg mentioned "something like the mean error", \cite{HU}). He did not find his relation via theoretical framework, he rather found it based on experimental estimations.

Few months later during the same year, and using the standard deviation of distribution of position and momentum, E.H.Kennard \cite{K} gave the theoretical framework that presented the uncertainty relation via the inequality
\begin{equation}
\sigma_x\sigma_p\geq{\hbar\over 2},
\end{equation}
where $\sigma_x$ and $\sigma_p$ represent a standard deviation of position and momentum obtained via repeated measurements. In the Heisenberg's paper of 1930 \cite{HEI1}, the uncertainty relation was given via the inequality $\delta p\ \delta q\gtrsim h$.

Since then, different works have been elaborated in the main objective to provide a rigorous proof of the uncertainty relation. We can site some of them \cite{PB, DO, FA, PR, SH} and the references list would be larger if we add the entropic uncertainty relation. It is true to say that the uncertainty principle was a surprising discovery of the twentieth centuries in physics that reinforced the divorce between classical point of view and quantum point of view regarding the possibility to associate simultaneously exact values of the momentum and position to any physical system.

Almost all works that provide a derivation of the uncertainty relation of position and momentum were made in the framework of quantum mechanics. In this paper we will provide a new approach to the uncertainty relation without using the quantum mechanics framework, and we will be able to elaborate the uncertainty relation with a precise mathematical description of the mean errors $\delta p$ and $\delta q$ and their simultaneous determination as well as interpretation.

The best candidate suitable to this purpose is the one described as "monster" or "pathology" by pioneers of classical analysis. The continuous and nowhere differentiable functions were described as monster because they mark a boundary between two different worlds: the set of functions used in classical physics (continuous and differentiable functions), and another world where the used functions are continuous but nowhere differentiable. These functions seem to be more like a treasure in our approach of the uncertainty relation.

\begin{figure}[h]\label{WE}
\begin{center}
\includegraphics[height=6cm, width=10cm]{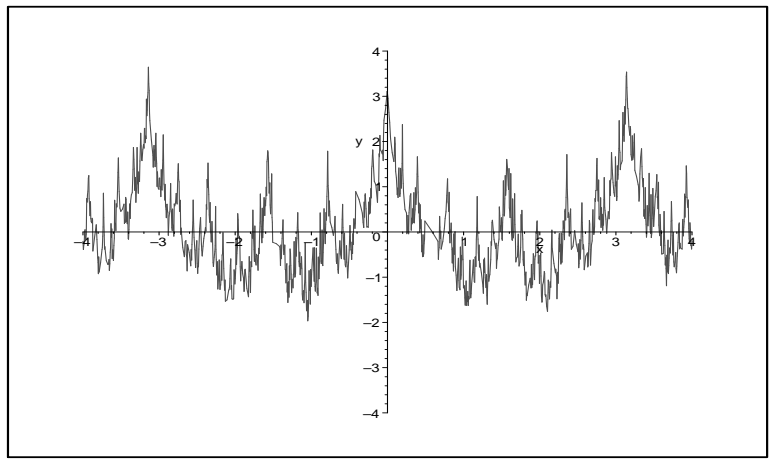}
\caption{\footnotesize{ Approximation of Weierstrass function
$W_\alpha(x)=\sum_{n=1}^{+\infty} q^{-\alpha n} \exp^{iq^{n}x}$,
$q=2$, $\alpha=0.3$.}}
\end{center}
\end{figure}

The first discovery of a continuous and nowhere differentiable function is thought to be due to B.Bolzano in 1830 (unpublished until 1890, \cite{BO}), and the first published function is due to K.Weierstrass (published in 1872, \cite{W, HA}). Here is an example of approximation of the graph of the Weierstrass function given in Fig.\ref{WE}.
\pesp

Usually the graph of this function is unreachable using a classical approach since it is impossible to determine a tangent line at any point (a non classical analysis based on fractional calculus, or fractal can be found in \cite{OLD,SKM,tri}). We can obtain a good approximation of the graph if we know the function and we can determine the error of approximation. However if any physical system has a position function which is continuous and nowhere differentiable, then we will prove that it is not possible to determine simultaneously the deviation of its position and the rate of change of this deviation: an uncertainty condition equivalent to the one discovered by Heisenberg imposes itself naturally that makes the position function unreachable.

This paper is organized as follow: in section 2, we introduce the needed analytical tools. The main results is introduced in section 3 and contains an introduction to the deviation of functions, the geometric and physical interpretation of the deviation as well as its normal rate of change, and a characterization of continuous and nowhere differentiable functions. At the end of this section, we prove that the nowhere differentiability leads straightforward to the obtention of an uncertainty relation in determining simultaneously the deviation and its normal rate of change, and we prove that this uncertainty relation is a fundamental characteristic of the nowhere differentiability of the used position function. In section 4, we present a general interpretation of the approach and a prospect for a new formalism that reflects the concept of uncertainty relation.

\section{Mean Functions}

Let $f$ be a continuous function defined on $[a,b]\subset\rR$. For all $x\in ]a,b[$ and for all $\varepsilon\in \rR^+$, we define the mean function by
\begin{equation}\label{F1}
f(x,\varepsilon)=\frac{1}{2\varepsilon}\int_{x-\varepsilon}^{x+\varepsilon}f(t)dt
\end{equation}
where the left and right mean functions by
\begin{equation}\label{F2}
f^-(x,\varepsilon)=\frac{1}{\varepsilon}\int_{x-\varepsilon}^{x}f(t)dt, \quad \quad f^+(x,\varepsilon)=\frac{1}{\varepsilon}\int_{x}^{x+\varepsilon}f(t)dt.
\end{equation}
We have for $\sigma=\pm$
\begin{equation}\label{F0}
f^\sigma(x,0)=\lim_{\varepsilon\rightarrow 0}f^\sigma(x,\varepsilon)=f(x),
\end{equation}
where the smaller $\varepsilon$ is, the closer the graph of $f^\sigma(x,\varepsilon)$ and the graph of $f$ are.\pesp

The definition of small resolution domain was introduced in \cite{BF}, as well as the following lemma, we will use them in the main result. Their proof comes straightforward from the following definition and a direct calculus of partial derivatives on the integral (\ref{F2}).

\begin{defn}\label{DR}
We call small resolution domain on $]a,b[$, and we denote it by ${\cal R}_f$ the set
\begin{equation}
{\cal R}_f= \Big\{\varepsilon\in \rR^+ \ / \ f^+(x,\varepsilon) \ \hbox{is differentiable on } ]a,b[\ \Big\}\cap [0,\alpha[,
\end{equation}
where $0<\alpha\ll 1$ is a small real number.
\end{defn}

\begin{lem}\label{L2} (\cite{BF})
The local small resolution domain depends on the differentiability of the function $f$ and we have

i) if $f$ is a nowhere differentiable function on $]a,b[$,  then ${\cal R}_{f}=]0,\alpha[$.

ii) if $f$ is a differentiable function on $]a,b[$, then ${\cal R}_{f}=[0,\alpha[ $.
\end{lem}

\begin{lem}\label{L1}(\cite{BF})
If $f$ is a continuous and nowhere differentiable function on $[a,b]$, then $\forall x\in [a,b]$, $\forall\varepsilon\in {\cal R}_f$, we have
\begin{equation}\label{F4}
f(x)= f^\sigma(x,\varepsilon) + \varepsilon\Big( \frac{\partial}{\partial \varepsilon}f^\sigma(x,\varepsilon)- \sigma\frac{\partial}{\partial x}f^\sigma(x,\varepsilon)\Big)\quad\hbox{for }
\sigma=\pm.
\end{equation}
\end{lem}
In the following we will only use $f^+(x,\varepsilon)$ as mean function of $f$ since $f^-(x+\varepsilon,\varepsilon)=f^+(x,\varepsilon)$.

\section{Main Results}

\subsection[Deviation and its Average Rate of Change]{Deviation of Function and its Average Rate of Change}

The lemma \ref{L1} has been proved for any nowhere differentiable function in \cite{BF}. Nevertheless the lemma \ref{L1} is still valid for any continuous function. Indeed, we have the following remark.

\begin{rem}\label{L1bis}
If $f\in C^0([a,b])$, then for all $x\in ]a,b[$, for all $\varepsilon\in {\cal R}_f$ we have
\begin{equation}\label{F5}
f(x)= f^+(x,\varepsilon) + \varepsilon\Big( \frac{\partial}{\partial \varepsilon}f^+(x,\varepsilon)- \frac{\partial}{\partial x}f^+(x,\varepsilon)\Big).
\end{equation}
\end{rem}

Indeed using (\ref{F2}) for $f\in C^0([a,b])$, we have
$\di \frac{\partial}{\partial \varepsilon}f^+(x,\varepsilon)=-\frac{1}{\varepsilon^2}\int_{x}^{x+\varepsilon}f(t)dt+\frac{1}{\varepsilon}f(x+\varepsilon)$, that is to say
\begin{equation}\label{F3}
\frac{\partial}{\partial \varepsilon}f^+(x,\varepsilon)=-\frac{1}{\varepsilon}f^+(x,\varepsilon)+\frac{1}{\varepsilon}f(x+\varepsilon).
\end{equation}
Moreover we have
\begin{equation}\label{F3Bis}
\frac{\partial}{\partial x}f^+(x,\varepsilon)=\frac{1}{\varepsilon}\Big(f(x+\varepsilon)-f(x)\Big)
\end{equation}
By substraction of (\ref{F3}) and (\ref{F3Bis}) we obtain
\begin{equation}
\varepsilon\Big( \frac{\partial}{\partial \varepsilon}f^+(x,\varepsilon)-\frac{\partial}{\partial x}f^+(x,\varepsilon)\Big)=f(x)-f^+(x,\varepsilon),
\end{equation}
which gives the result.

\rightline\Box

{\bf Quantum Resolution Domain}\pesp

The Definition \ref{DR} gives: if $0\notin{\cal R}_f$, then the function $f$ is continuous and non differentiable on $]a,b[$, where $f$ can be nowhere differentiable on $]a,b[$, or almost everywhere non differentiable with points of differentiability on $[a,b]$, or differentiable with points of non differentiability on $[a,b]$.\pesp

Since we want to provide a deterministic understanding of the Heisenberg's uncertainty relation, the set of continuous functions that will be concerned by Definition \ref{DR} if $0\not\in{\cal R}_f$ is the set of continuous and nowhere differentiable functions on $]a,b[$. Indeed, the set of continuous functions that are almost everywhere non differentiable with points of differentiability on $[a,b]$, and the set of continuous functions that are differentiable with points of non differentiability on $[a,b]$ have to be excluded from this approach, because if a quantum physical system has an unreachable trajectory defined by a continuous position function given by $S=f(t)$ (a physical system with unreachable continuous position function means that the physical system is moving in a continuous medium, however the motion of the physical system is completely unknown), then it is impossible to determine simultaneously at a given $t_0$ its position $f(t_0)$ and its velocity $f'(t_0)$.\pesp

Taking into account the previous comment, we define the restriction ${\cal Q}_f$ of the set ${\cal R}_f$ as follow.

\begin{defn}\label{DR1}
Let $f\in C^0([a,b])$.
We call quantum resolution domain on $]a,b[$, and we denote it by ${\cal Q}_f$ the set
\begin{equation}
{\cal Q}_f= \Big\{\varepsilon\in  [0,\alpha[ \ / \ f^+(x,\varepsilon) \ \hbox{is differentiable on } ]a,b[\ \Big\},\qquad \hbox{where}\quad  0<\alpha\ll 1
\end{equation}
that satisfies: if $0\notin{\cal Q}_f$, then $f$ is nowhere differentiable on $]a,b[$.
\end{defn}

\begin{rem}
 We always have: if $0\in{\cal Q}_f$, then $f$ is differentiable on $]a,b[$.
\end{rem}

\begin{defn}
Let $f\in C^0([a,b])$. For all $x\in ]a,b[$ and for all $\varepsilon\in{\cal Q}_f$, we call deviation of the function $f^+(x,\varepsilon)$ on $[0,\varepsilon]$ and we denote $\Delta_\varepsilon f(x)$ the quantity
\begin{equation}\label{D1}
\Delta_\varepsilon f(x)=f^+(x,0)- f^+(x,\varepsilon).
\end{equation}
\end{defn}

\begin{prop}\label{P1}
Let $f\in C^0([a,b])$ be a nowhere differentiable function on $]a,b[$, then we have for all $\varepsilon\in{\cal Q}_f$ and for all $x\in ]a,b[$
\begin{equation}\label{Dv}
\Big( \frac{\partial}{\partial \varepsilon}f^+(x,\varepsilon)-\frac{\partial}{\partial x}f^+(x,\varepsilon)\Big)= \di\frac{\Delta_\varepsilon f(x)}{\varepsilon}.
\end{equation}
\end{prop}

{\it Proof:} Since $f$ is nowhere differentiable on $]a,b[$, then ${\cal Q}_f=]0,\alpha[$. Using Lemma \ref{L1bis}, we can divide the formula (\ref{F5}) by $\varepsilon\neq0$ and we obtain the result.

\rightline\Box

\begin{defn}\label{Def0}
Let $f\in C^0([a,b])$. For all $\varepsilon>0$ and for all $x\in]a,b[$, we call
average rate of change of the $f^+(x,\varepsilon)$ on $[0,\varepsilon]$ the quotient
\begin{equation}
\Delta_\varepsilon V (x)=\di-\frac{\Delta_\varepsilon f(x)}{\varepsilon}.
\end{equation}
and for $\Delta_\varepsilon V (x)\neq0$, we denote
\begin{equation}
\Delta_\varepsilon P (x)=\di-\frac{1}{\Delta_\varepsilon V (x)}
\end{equation}

\end{defn}

\subsection{Interpretation }
\subsubsection{Average Rate of Change }
Let us consider
\begin{equation}\label{S}
{\cal S}=\Big\{(x,\varepsilon,z)\in ]a,b[\times{\cal Q}_f\times\rR\ /\ z=f^+(x,\varepsilon)\Big\}
\end{equation}
that represents a surface in $\rR^3$ and

\begin{equation}
{\cal S_{x_0}}={\cal S}\cap\Big\{(x,\varepsilon,z)\in \rR^3\ /\ x=x_0\Big\}
\end{equation}
that represents the curve described by $z=f^+(x_0,\varepsilon)$ for all $\varepsilon\in{\cal Q}_f$, and let us consider

\begin{equation}
{\cal S_{\varepsilon_0}}={\cal S}\cap\Big\{(x,\varepsilon,z)\in \rR^3\ /\ \varepsilon=\varepsilon_0\Big\}
\end{equation}
that represents the curve described by $z=f^+(x,\varepsilon_0)$ for all $x\in ]a,b[$.

\pesp

Since $f(x)=f^+(x,0)$ by (\ref{F0}), then for all $\varepsilon\in{\cal Q}_f$, $\di \Delta_\varepsilon V (x)=\frac{f^+(x,\varepsilon)-f^+(x,0)}{\varepsilon}$, thus $\Delta_\varepsilon V (x)=\di -\frac{\Delta_\varepsilon f(x_0)}{\varepsilon}$ can be geometrically interpreted as the average rate of change of $f^+(x,\varepsilon)$ with respect to $\varepsilon$ in the interval $[0,\varepsilon]$, which represents the slope of the secant line $\cal L_s$ to the graph of ${\cal S_{x_0}}$ through the points $P_1(x_0,\varepsilon,f^+(x_0,\varepsilon))$ and $P_2(x_0,0,f^+(x_0,0))$.
The quantity $\Delta_\varepsilon P (x)=\di -\frac{1}{\Delta V_\varepsilon (x_0)}=\di\frac{\varepsilon}{\Delta_\varepsilon f(x_0)}$ geometrically represents the average rate change of $f^+(x,\varepsilon)$ along a line $\cal L$ that is perpendicular to the line $\cal L_s$ since $\Delta V_\varepsilon (x_0)\Delta_\varepsilon P (x_0)=-1$.
Along the line $\cal L$, a change of one unit in $\varepsilon$ corresponds to a change of $\di \frac{\varepsilon}{\Delta_\varepsilon f(x_0)}$ in $z=f(x_0,\varepsilon)$.
We call $\Delta_\varepsilon P (x)$ the normal average rate change of $f^+(x,\varepsilon)$ to the line $\cal L_s$.\pesp

\subsubsection{Physical Interpretation of $S=f^+(t,\varepsilon)$}

Based on the previous interpretations we have:

\ni i) The quantity $f^+(x,\varepsilon)$ represents the approximation of the position function $f(x)$ at $\varepsilon\in{\cal Q}_f$.

\ni ii) The quantity $\Delta_\varepsilon f(x)$ represents the error of the approximation of the position function $f$.

\ni iii) The quantity $\Delta_\varepsilon P (x)$ represents the average rate change of the approximation of the position function $f(x)$ along the line $\cal L$ that is perpendicular to the line $\cal L_s$.\pesp

The motion of a physical system is completely known if the physical system's position in space is known at all time as well as the time derivative of this position function. However if the physical system's position in space is known at all time meanwhile it is impossible to determine the time derivative of this position function, the motion of the physical system will be unknown. Then if we consider a physical system that has a trajectory defined by $S=f(t)$ a continuous and nowhere differentiable function, the motion of this physical system is unknown, meanwhile the motion of the physical system that has a trajectory defined by $S'=f^+(t,\varepsilon)$ is known at all time for any positive $\varepsilon$, where the position function $S'=f^+(t,\varepsilon)$ for a given $\varepsilon>0$ represents the approximation of the unknown motion of the physical system that has a trajectory defined by $S=f(t)$.\pesp

Following the previous comment, $\Delta_\varepsilon f(t)$ corresponds to the uncertainty in measuring the deviation of the known motion of the physical system  that has a trajectory defined by $S'=f^+(t,\varepsilon)$ with respect to the unknown motion of the physical system that has a trajectory defined by $S=f(t)$ on the interval $[0,\varepsilon]$, and we have:
\begin{itemize}
  \item If $\varepsilon=\varepsilon_0$, $f^+(t,\varepsilon_0)$ represents the trajectory of a physical system that one can observe or measure (observable position function).
  \item If $t=t_0$, $f^+(t_0,\varepsilon)$ represents all possible positions of a physical system that can be observed at $t_0$.
  \item If $t=t_0$ and $\varepsilon=\varepsilon_0$, $f^+(t_0,\varepsilon_0)$ represents the observable position of a physical system at the point $(t_0,\varepsilon_0)$.
  \item If $(t,\varepsilon)\in ]a,b[\times{\cal Q}_f$, then $f^+(t,\varepsilon)$ represents all possible differentiable trajectories (observable positions function)  of the physical system.
  \item If $\varepsilon=0$, the trajectory of a physical system $S=f^+(t,0)=f(t)$ is unreachable.
\end{itemize}
Thus $\Delta_\varepsilon P (t)$, for a positive $\varepsilon$, corresponds to the uncertainty in measuring the average velocity of the deviation of the known motion of the physical system  that has a trajectory defined by $S'=f^+(t,y)$ for $y\in[0,\varepsilon]$ along the line $\cal L$ that is perpendicular to the line $\cal L_s$.
\pesp

Some characteristics of a physical system with unknown motion given by a trajectory defined by $S=f(t)$ can be obtained via characteristics of the approximated motion given by $S'=f^+(t,\varepsilon)$ for all $(t,\varepsilon) \in I\times {\cal Q}_f$. In particular the deviation of the known motion from the unknown one and the average rate of change of this deviation is subject to an uncertainty relation that constitutes the central point of the following formalism.

\subsection[Characterization of Nowhere Differentiability]{Characterization of Nowhere Differentiable Function}

\begin{lem}\label{L3}
Let $f\in C^0([a,b])$.
\begin{equation}
f \ \hbox{is differentiable on}\  ]a,b[\quad \Longleftrightarrow \quad \exists\varepsilon\in {\cal Q}_f\ / \quad\forall x\in]a,b[\quad \Delta_\varepsilon f(x)=0.
\end{equation}
\end{lem}

{\it Proof:}
If $f$ is differentiable on $]a,b[$, then ${\cal Q}_f=[0,\alpha[$ by Lemma \ref{L2} and Definition \ref{DR1}. For $\varepsilon=0\in{\cal Q}_f$, we have $f(x)=f^+(x,0)=\di \lim_{\varepsilon\rightarrow 0}f^+(x,\varepsilon)$ for all $x\in]a,b[$, then there exists $\varepsilon\in {\cal Q}_f$ such that $\Delta_\varepsilon f(x)=0$ for all $x\in]a,b[$.

Inversely, if there exists $\varepsilon_0\in{\cal Q}_f$ such that for all $x\in]a,b[$, $\Delta_{\varepsilon_0} f(x)=0$, then we have two cases: $\varepsilon_0=0$ or $\varepsilon_0\neq0$. If $\varepsilon_0=0\in{\cal Q}_f$, then for all $x\in]a,b[$ we have $f(x)=f^+(x,0)$ which gives $f(x)$ differentiable by the Definition \ref{DR1}. If $\varepsilon_0\neq0$, then
by (\ref{D1}) we have $f(x)=f^+(x,\varepsilon_0)$ for all $x\in]a,b[$, which implies that
$f'(x)=\di\frac{1}{\varepsilon_0}\Big(f(x+\varepsilon_0)-f(x)\Big)$ for all $x\in]a,b[$. Therefore the function $f$ is differentiable on $]a,b[$, which concludes the proof.

\rightline\Box

\begin{cor}\label{Cor1}
Let $f\in C^0([a,b])$.
\begin{equation}
f\ \hbox{nowhere differentiable on}\ ]a,b[\quad \Longleftrightarrow
\quad \forall \varepsilon\in {\cal Q}_f,\quad \exists\ x\in]a,b[\ /\quad \Delta_{\varepsilon} f (x)\neq0.
\end{equation}
\end{cor}

{\it Proof:} By contraposition, let us suppose that there exists $\varepsilon\in {\cal Q}_f$ such that for all $\ x\in]a,b[$, we have $\Delta_{\varepsilon} f (x)=0$,
then by lemma \ref{L3}, $f$ is differentiable on $]a,b[$.

Inversely, if for all $\varepsilon\in {\cal Q}_f$, there exists $x\in]a,b[$ such that $\Delta_{\varepsilon} f (x)\neq0,$
then we have two cases: $0\in {\cal Q}_f\ $ or $\ 0\not\in {\cal Q}_f$. Let us suppose that $0\in {\cal Q}_f$, then there exists $x_0\in ]a,b[$ such that $\Delta_{0} f (x_0)\neq0$, that gives $f(x_0)\neq f^+(x_0,0)$, which is a contradiction. Therefore $0\not\in{\cal Q}_f$, and by Definition \ref{DR1}, we obtain that $f$ is nowhere differentiable on $]a,b[$.

\rightline\Box

\begin{defn}
Let $f\in C^0([a,b])$. For all $\varepsilon\in{\cal Q}_f$, we define
\begin{equation}\label{Is}
{\cal I}_\varepsilon=\Big\{\ x\in ]a,b[\ /\ \Delta_\varepsilon f(x)=0\ \Big\}.
\end{equation}
\end{defn}

\begin{prop}\label{Pr1} Let $f\in C^0([a,b])$.
The function $f$ is nowhere differentiable on $]a,b[$ if and only if for all $\varepsilon\in{\cal Q}_f$ and for all $x\in{\cal I}_\varepsilon$, there does not exist any neighborhood $V_x$ of $x$ included in ${\cal I}_\varepsilon$.
\end{prop}

{\it Proof:} Let $f\in C^0([a,b])$ nowhere differentiable on $]a,b[$. Let us suppose that there exists $\varepsilon_0\in{\cal Q}_f$, there exists $x_0\in{\cal I}_{\varepsilon_0}$ such that we can find a neighborhood $ V_{x_0}$ of $x_0$ included in ${\cal I}_{\varepsilon_0}$. Then for all $x$ in $V_{x_0}$ we have $f(x)=f^+(x,\varepsilon_0)$, and thus $f(x)$ is differentiable on $V_{x_0}$, which is a contradiction.

Inversely, let us suppose that for all $\varepsilon\in{\cal Q}_f$, for all  $x\in{\cal I}_\varepsilon $, there does not exist any neighborhood $\ V_x$ of $x$ included in ${\cal I}_\varepsilon$. Then $\forall x\in{\cal I}_\varepsilon $, for all neighborhood $\ V_x\subset]a,b[$, there exists $y\in \ V_x$ such that $y\not\in{\cal I}_{\varepsilon}$, then $\Delta_{\varepsilon} f(y)\not=0$, therefore: $\forall\varepsilon\in{\cal Q}_f$, there exists  $y\in]a,b[$ such that $\Delta_{\varepsilon} f(y)\not=0$, then by Corollary \ref{Cor1}, we obtain that $f$ is nowhere differentiable.

\rightline\Box

\subsection{Uncertainty Principle}

We would like to express the uncertainty theorem in the spirit of Heisenberg as follow:
\ppesp
{\bf Uncertainty Principle}
{\it Let $f\in C^0([a,b])$. If $f$ is a nowhere differentiable function on $]a,b[$,
then it is impossible to determine simultaneously the deviation $\Delta_\varepsilon f(x)$ and the normal average rate of change  $\Delta_\varepsilon P(x)$ to an accuracy equal to $\Delta_\varepsilon f(x)\ \Delta_\varepsilon P(x)= \varepsilon$.}\ppesp

However the statement of the previous theorem does not provide a clear precise mathematical formulation of the used concepts. The Heisenberg's uncertainty can be provided by the following theorem.

\begin{thm} (Uncertainty Principle)\label{UP}
Let $f\in C^0([a,b])$. If $f$ is nowhere differentiable on $]a,b[$,
then for all $\varepsilon\in{\cal Q}_f$, for all $x\in]a,b[\ \backslash{\cal I}_{\varepsilon}$, the quantities $\Delta_{\varepsilon} f(x)$ and $\Delta_{\varepsilon} P(x)$ are well  defined to an accuracy equal to
\begin{equation}\label{IP}
\Delta_{\varepsilon} f(x)\ \Delta_{\varepsilon} P(x)= \varepsilon
\end{equation}
\end{thm}

{\it Proof:} The proof is straight forward from equality (\ref{F5}). Indeed, for $f\in C^0([a,b])$ we have always
\begin{equation}
\Delta_\varepsilon f(x)=\varepsilon \Big(\frac{\partial}{\partial \varepsilon}f^+(x,\varepsilon)-\frac{\partial}{\partial x}f^+(x,\varepsilon)\Big).
\end{equation}
Since  $f$ is nowhere differentiable on $]a,b[$, then by Corollary \ref{Cor1}, $\forall \varepsilon\in {\cal Q}_f$, $\exists\ x\in]a,b[$ such that $\Delta_{\varepsilon} f (x)\neq0$, and then $\forall \varepsilon\in {\cal Q}_f$, $\exists\ x\in]a,b[$ such that $\Big(\frac{\partial}{\partial \varepsilon}f^+(x,\varepsilon)-\frac{\partial}{\partial x}f^+(x,\varepsilon)\Big)\neq0$, which gives that for all $\varepsilon\in{\cal Q}_f$, for all $x\in]a,b[\ \backslash{\cal I}_{\varepsilon}$ we have (\ref{IP}) by Definition \ref{Def0}.

\rightline\Box
\pesp

The Theorem \ref{UP} of the Uncertainty Principle is in fact equivalent to say that if $f$ is continuous and nowhere differentiable, then one can not obtain by any arbitrary measurement (for a fixed $\varepsilon$ or a fixed $x$ or both of them) the quantities  $\Delta_{\varepsilon} f(x)$ and $\Delta_{\varepsilon} P(x)$ to an accuracy equal to (\ref{IP}).
To express the limitation to determine simultaneously the deviation and the normal rate of change of this deviation (as the limitation imposed in the quantum mechanics), we can express the uncertainty principle as follow.

\begin{cor} (Uncertainty Principle) Let $f\in C^0([a,b])$.
If the function $f$ is nowhere differentiable on $]a,b[$, then it is impossible to have the following situations:

i) There exists $\varepsilon_0\in{\cal Q}_f$ such that for all $x\in]a,b[\ \backslash{\cal I}_{\varepsilon_0}$, $\Delta_{\varepsilon_0} f(x)$ and $\Delta_{\varepsilon_0} P(x)$ are well defined to an accuracy equal to
\begin{equation}\label{IP01}
\Delta_{\varepsilon_0} f(x)\ \Delta_{\varepsilon_0} P(x)= \varepsilon_0.
\end{equation}

ii) There exists $x_0\in\ ]a,b[\ \backslash \cup_{\varepsilon\in{\cal Q}_f}{\cal I}_{\varepsilon}$ such that for all $\varepsilon\in{\cal Q}_f$, $\Delta_\varepsilon f(x_0)$ and $\Delta_\varepsilon P(x_0)$ are well defined to an accuracy equal to
\begin{equation}\label{IP02}
\Delta_\varepsilon f(x_0)\ \Delta_\varepsilon P(x_0)= \varepsilon.
\end{equation}

iii) There exists $\varepsilon_0\in{\cal Q}_f$,  there exists $x_0\in]a,b[\ \backslash{\cal I}_{\varepsilon_0}$ such that $\Delta_{\varepsilon_0} f(x_0)$ and $\Delta_{\varepsilon_0} P(x_0)$ are well defined to an accuracy equal to
\begin{equation}\label{IP03}
\Delta_{\varepsilon_0} f(x_0)\ \Delta_{\varepsilon_0} P(x_0)= \varepsilon_0.
\end{equation}

\end{cor}

{\it Proof:} i) Let $f\in C^0([a,b])$ be a nowhere differentiable function on $]a,b[$.
By contradiction let us suppose that there exists $\varepsilon_0\in {\cal Q}_f$ such that for all $x\in ]a,b[\ \backslash{\cal I}_{\varepsilon_0}$ we can determine simultaneously the deviation $\Delta_{\varepsilon_0} f(x)$ and the normal average rate of change  $\Delta_{\varepsilon_0} P(x)$ to an accuracy equal to (\ref{IP01}). Then we have (\ref{IP01}) for $\varepsilon=\varepsilon_0$ with
\begin{equation}
\left\{
  \begin{array}{ll}
    \Delta_{\varepsilon_0} f(x)= & f(x)-f^+(x,\varepsilon_0) \\
    \Delta_{\varepsilon_0} P(x)= &\di {\varepsilon_0 \over\Delta_{\varepsilon_0} f(x)}=\Big(\frac{\partial}{\partial \varepsilon}f^+(x,\varepsilon_0)-\frac{\partial}{\partial x}f^+(x,\varepsilon_0)\Big)^{-1}
  \end{array}
\right.
\end{equation}
which means that the deviation $\Delta_{\varepsilon_0} f(x)$ is constant with respect to $\varepsilon$ and function of $x$. Therefore we have
$\di\frac{\partial}{\partial \varepsilon}f^+(x,\varepsilon_0)=0$, which gives
\begin{equation}
\left\{
  \begin{array}{ll}
    \Delta_{\varepsilon_0} f(x)= & f(x)-f^+(x,\varepsilon_0) \\
    \Delta_{\varepsilon_0} P(x)=&\di {\varepsilon_0 \over\Delta_{\varepsilon_0} f(x)}=-\Big(\frac{\partial}{\partial x}f^+(x,\varepsilon_0)\Big)^{-1}
  \end{array}
\right.
\end{equation}
 for all $x\in]a,b[\ \backslash{\cal I}_{\varepsilon_0}$. Then we have
\begin{equation}
\frac{\Delta_{\varepsilon_0} f(x)}{\varepsilon_0}=-\frac{\partial}{\partial x}f^+(x,\varepsilon_0)=-\frac{1}{\varepsilon_0}\Big(f(x+\varepsilon_0)-f(x)\Big).
\end{equation}
Therefore
\begin{equation}
\frac{f(x)-f^+(x,\varepsilon_0)}{\varepsilon_0}=-\frac{1}{\varepsilon_0}\Big(f(x+\varepsilon_0)-f(x)\Big),
\end{equation}
and then for all $x\in]a,b[\ \backslash{\cal I}_{\varepsilon_0}$
\begin{equation}\label{F6}
f(x+\varepsilon_0)=f^+(x,\varepsilon_0).
\end{equation}
Then by Proposition \ref{Pr1} for all $x\in]a,b[\ \backslash{\cal I}_{\varepsilon_0}$, there exists an open neighborhood $V_x$ of $x$ on which the equality (\ref{F6}) holds good. This implies that
$f(x+\varepsilon_0)$ is differentiable on $V_x$, which is impossible. Thus we can not simultaneously determine the deviation $\Delta_{\varepsilon_0} f(x)$ and the normal average rate of change  $\Delta_{\varepsilon_0} P(x)$ on $[0,\varepsilon_0]$ to an accuracy equal to (\ref{IP01}).

ii) Let $f\in C^0([a,b])$ be a nowhere differentiable function on $]a,b[$.
By contradiction let us suppose that there exists $x_0\in ]a,b[\ \backslash \cup_{\varepsilon\in{\cal Q}_f}{\cal I}_{\varepsilon}$ such that for all $\varepsilon\in {\cal Q}_f$ we can simultaneously determine the deviation $\Delta_{\varepsilon} f(x_0)$ and the normal average rate of change $\Delta_{\varepsilon} P(x_0)$ for the same $x_0$ to an accuracy equal to (\ref{IP02}). Then we have (\ref{IP02}) for $x=x_0$ with
\begin{equation}
\left\{
  \begin{array}{ll}
    \Delta_{\varepsilon} f(x_0)= & f(x_0)-f^+(x_0,\varepsilon) \\
    \Delta_{\varepsilon} P(x_0)= &\di {\varepsilon \over\Delta_{\varepsilon} f(x_0)}=\Big(\frac{\partial}{\partial \varepsilon}f^+(x_0,\varepsilon)-\frac{\partial}{\partial x}f^+(x_0,\varepsilon)\Big)^{-1}
  \end{array}
\right.
\end{equation}
which means that the deviation $\Delta_{\varepsilon} f(x_0)$ is constant with respect to $x$ and function of $\varepsilon$ as well as $\Delta_{\varepsilon} P(x_0)$, then we have
$\di\frac{\partial}{\partial x}f^+(x_0,\varepsilon)=0$, which gives for all $\varepsilon\in{\cal Q}_f$:
\begin{equation}
\left\{
  \begin{array}{ll}
    \Delta_{\varepsilon} f(x_0)= & f(x_0)-f^+(x_0,\varepsilon) \\
    \Delta_{\varepsilon} P(x_0)=&\di {\varepsilon \over\Delta_{\varepsilon} f(x_0)}=\Big(\frac{\partial}{\partial \varepsilon}f^+(x_0,\varepsilon)\Big)^{-1}.
  \end{array}
\right.
\end{equation}
Then we have
\begin{equation}
\frac{\Delta_{\varepsilon} f(x_0)}{\varepsilon}=\Big(\frac{\partial}{\partial \varepsilon}f^+(x_0,\varepsilon)\Big)=-\frac{1}{\varepsilon}f^+(x_0,\varepsilon)+\frac{1}{\varepsilon}f(x_0+\varepsilon).
\end{equation}
Therefore
\begin{equation}
\frac{f(x_0)-f^+(x_0,\varepsilon)}{\varepsilon}=-\frac{1}{\varepsilon}f^+(x_0,\varepsilon)+\frac{1}{\varepsilon}f(x_0+\varepsilon),
\end{equation}
and then there exists $x_0\in ]a,b[\ \backslash \cup_{\varepsilon\in{\cal Q}_f}{\cal I}_{\varepsilon}$ such that for all $\varepsilon\in{\cal Q}_f$
\begin{equation}\label{F9}
f(x_0)=f(x_0+\varepsilon).
\end{equation}
which means that there exists an interval $[x_0,x_0+\alpha[\subset ]a,b[$, where $\alpha$ is given by Definition \ref{DR1}, on which the function $f$ is constant and then differentiable, which is impossible since $f$ is nowhere differentiable on $]a,b[$.

iii) Let $f\in C^0([a,b])$ be a nowhere differentiable function on $]a,b[$.
By contradiction let us suppose that there exists $\varepsilon_0\in{\cal Q}_f$, there exists $x_0\in]a,b[\ \backslash{\cal I}_{\varepsilon_0}$ such that we can simultaneously determine $\Delta_{\varepsilon_0} f(x_0)$ and $\Delta_{\varepsilon_0} P(x_0)$ to an accuracy equal to (\ref{IP03}). Then we have (\ref{IP03}) for $x=x_0$ and $\varepsilon=\varepsilon_0$ with
\begin{equation}
\left\{
  \begin{array}{ll}
    \Delta_{\varepsilon_0} f(x_0)= & f(x_0)-f^+(x_0,\varepsilon_0) \\
    \Delta_{\varepsilon_0} P(x_0)= &\di {\varepsilon_0 \over\Delta_{\varepsilon_0} f(x_0)}=\Big(\frac{\partial}{\partial \varepsilon}f^+(x_0,\varepsilon_0)-\frac{\partial}{\partial x}f^+(x_0,\varepsilon_0)\Big)^{-1}
  \end{array}
\right.
\end{equation}
Since $f^+(x_0,\varepsilon_0)$ is constant with respect to $x$ and $\varepsilon$, then
\begin{equation}\label{F10}
\di\frac{\partial}{\partial x}f^+(x_0,\varepsilon_0)=  0 \quad \hbox{and}\quad
\di\frac{\partial}{\partial \varepsilon}f^+(x_0,\varepsilon_0)=0
  \end{equation}
Using (\ref{F10}) and Proposition \ref{P1}, we have
\begin{equation}
\left\{
  \begin{array}{ll}
    \Delta_{\varepsilon_0} f(x_0)= & 0 \\
    \Delta_{\varepsilon_0} P(x_0) & \hbox{not defined}
  \end{array}
\right.
\end{equation}
which is impossible since $x_0\not\in{\cal I}_{\varepsilon_0}$ and ends the proof of iii).

\rightline\Box

\begin{prop}
Let $f\in C^0([a,b])$. If $f$ is nowhere differentiable on $]a,b[$, then for all $\varepsilon\in{\cal Q}_f$, for all $x\in ]a,b[\ \backslash{\cal I}_\varepsilon$ both uncertainties $\Delta_\varepsilon f(x)$ and $\Delta_\varepsilon P (x)$ cannot be simultaneously small.
\end{prop}

{\it Proof:}
From the continuity of $f$ on $[a,b]$, for all $\delta>0$ there exists $\eta(x,\delta)>0$ such that for $\vert x-t\vert<\eta$ we have $\vert f(x)-f(t)\vert<\delta$.
We deduce that
\begin{equation}\label{E2}
\forall \varepsilon<\eta,\qquad\vert{1\over \varepsilon}\int^{x+\varepsilon}_x (f(x)-f(t)) dt\vert<\delta
\end{equation}
that is to say
\begin{equation}\label{E1}
\vert\Delta_\varepsilon f(x)\vert<\delta.
\end{equation}
Since $f$ is continuous on the closed bounded interval $[a,b]$, then $f$ is uniformly continuous and we can consider that $\eta$ does not depend on $x$. Let us suppose now that both uncertainties $\Delta_\varepsilon f(x)$ and $\Delta_\varepsilon P (x)$ can be simultaneously small to an accuracy equal to (\ref{IP}),
then for all $\varepsilon\in {\cal Q}_f$ with $\varepsilon<\eta$, for all $x\in]a,b[\backslash I_{\varepsilon}$, we have $\Delta_{\varepsilon} f(x)$ small as well as $\Delta_{\varepsilon} P (x)$.
The inequalities (\ref{E1}) and (\ref{IP}) give for $\varepsilon\in{\cal Q}_f$ and for $x\in]a,b[\ \backslash I_{\varepsilon}$
\begin{equation}
\vert\Delta_{\varepsilon} f(x)\vert\ \vert\Delta_{\varepsilon}P(x)\vert=\varepsilon<\delta\ \vert\Delta_{\varepsilon}P(x)\vert,
\end{equation}
then we have
\begin{equation}
\vert\Delta_{\varepsilon}P(x)\vert>{\varepsilon\over \delta }.
\end{equation}
In particular for $\di\delta={1\over n}$, we have
\begin{equation}
\left\{
  \begin{array}{ll}
    \vert\Delta_{\varepsilon} P(x)\vert & > n\varepsilon \\
    \vert\Delta_{\varepsilon} f(x)\vert & < \di{1\over n}
  \end{array}
\right.
\end{equation}
then for $n$ big enough we obtain simultaneously $\Delta_\varepsilon f(x)$ small and $\Delta_\varepsilon P (x)$ big, which is a contradiction.

\rightline\Box

\subsection{General Interpretation of the Uncertainty Principle}

To give an adequate interpretation that raises the physical concept conveyed in the previous framework we need to introduce the following definition.
\begin{defn} Let $f\in C^0([a,b])$, and let $\Gamma_f=\{(x,f(x))\in\rR^2\ /\ x\in[a,b]\}$ be the graph of $f$.
We say that $\Gamma_f$ is observable if $f$ is differentiable on $]a,b[$ and we say that $\Gamma_f$ is non observable if $f$ is nowhere differentiable on $]a,b[$.
\end{defn}

\begin{rem}
The observable graph $\Gamma_f$ is the one that can be drawn with our tools, that is to say its location in a given reference frame and its instantaneous rate of change can be determined for all $x\in]a,b[$, which leads to obtain an exact shape of the graph $\Gamma_f$ and the precise location of its key features everywhere. However the non observable graph $\Gamma_f$ is the one that can not be drawn in a given reference frame, and an exact shape as well as a precise location of its key features cannot be obtained anywhere, that is to say it is impossible to determine its instantaneous rate of change for all $x\in]a,b[$. The non observable graph $\Gamma_f$ is unreachable in a given reference frame, nevertheless the graph of the mean of $f$ given by (\ref{F1}) is always observable.
\end{rem}

Apparently, the framework of the formalism that provides analysis and accurate description of the nowhere differentiable functions differs radically from the one that provides analysis and description of the differentiable functions in a deep sense other than differentiability and non differentiability, indeed:

1) From the Remark \ref{L1bis} we can understand that it is impossible to obtain an accurate description of the deviation $\Delta_{\varepsilon} f(x)$ from  $\di\varepsilon\Big( \frac{\partial}{\partial \varepsilon}f^+(x,\varepsilon)- \frac{\partial}{\partial x}f^+(x,\varepsilon)\Big)$ if we fix one of the variables or both of them in $f^+(x,\varepsilon)$. The non observable graph $\Gamma_f$ of the nowhere differentiable function $f$ cannot be well described by considering only one observable graph
\begin{equation}
\Gamma_{f_{\varepsilon_0}^+}=\Big\{(x,f^+(x,\varepsilon_0))\in\rR^2\ /\ x\in[a,b]\Big\},
\end{equation}
as the best representant of $\Gamma_f$. Indeed the equality (\ref{F5}) becomes false if we fix one of the parameters $\varepsilon_0$ or $x_0$, or both of them before to elaborate the partial derivatives with respect to $\varepsilon$ or with respect to $x$.

2) If any physical system has a position function $s=f(t)$ which is continuous and nowhere differentiable, then it is impossible to assign to the physical system exact simultaneous small values to the deviation of its position $\Delta_\varepsilon f(t)$ and to the normal average rate of change $\Delta_{\varepsilon} P(t)$ to an accuracy equal to (\ref{IP}). Indeed, if we want to measure the deviation of its position $\Delta_\varepsilon f(t)$ or its normal average rate of change $\Delta_{\varepsilon} P(t)$ we will be obliged to fix one of the two variables that define the surface of all possible approximations of $f$, and then the equality (\ref{IP})
becomes impossible to obtain. Nevertheless these quantities can be determined with some uncertainty.

3) The uncertainty principle says that if $f$ is a nowhere differentiable function on $]a,b[$, then it is impossible to determine simultaneously the deviation $\Delta_\varepsilon f(x)$ and the normal average rate of change  $\Delta_\varepsilon P(x)$ on $[0,\varepsilon]$ to an accuracy equal to (\ref{IP}), and if $\varepsilon$ tends to 0, the deviation $\Delta_\varepsilon f(x)$ will tend to 0 and the normal average rate of change  $\Delta_\varepsilon P(x)$ will be not defined.
That is why the uncertainty inequality can be obtained if we admit, as in the quantum physics framework, that the choice of $\varepsilon$ is subject to some natural physical limitation and that it can never be smaller than a fixed fraction of Plank's constant. This physical limitation will be characterized by the minimum value of $\varepsilon$ that makes the determination of the deviation $\Delta_\varepsilon f(x)$ as well as the normal average rate of change  $\Delta_\varepsilon P(x)$ well defined to an accuracy equal to (\ref{IP}).

\subsection{The Converse of the Uncertainty Principle}

Let us consider a physical system that has a continuous position function $f$ (that is to say the system is moving in a continuous medium). If it is impossible to simultaneously determine the deviation $\Delta_{\varepsilon} f(x)$ and the normal average rate of change $\Delta_{\varepsilon} P(x)$ on $[0,\varepsilon]$ to an accuracy equal to $(\ref{IP})$, is it possible to deduce some information concerning the regularity of the position function. The converse of the Uncertainty Principle answers this question as follow.

\begin{thm}\label{IUP} (Converse Uncertainty Principle)
Let $f\in C^0([a,b])$.
If for all $\varepsilon\in{\cal Q}_f$, for all $x\in]a,b[\ \backslash{\cal I}_{\varepsilon}$ the quantities $\Delta_{\varepsilon} f(x)$ and $\Delta_{\varepsilon} P(x)$ are well defined to an accuracy equal to
\begin{equation}\label{IP2}
\Delta_{\varepsilon} f(x)\ \Delta_{\varepsilon} P(x)= \varepsilon,
\end{equation}
then $f$ is nowhere differentiable on $]a,b[$.
\end{thm}

{\it Proof:} Let us suppose that for all $\varepsilon\in{\cal Q}_f$, for all $x\in]a,b[\backslash{\cal I}_{\varepsilon}$ the quantities $\Delta_{\varepsilon} f(x)$ and $\Delta_{\varepsilon} P(x)$ are well defined to an accuracy equal to (\ref{IP2}).
If for all $\varepsilon\in{\cal Q}_f$ we have $]a,b[\ \backslash{\cal I}_{\varepsilon}=\emptyset$, then we obtain that for all $\varepsilon\in{\cal Q}_f$, there is no $x\in]a,b[\backslash{\cal I}_{\varepsilon}$, such that $\Delta_{\varepsilon} f(x)$ and $\Delta_{\varepsilon} P(x)$ are well defined to an accuracy equal to (\ref{IP2}), which is excluded. Thus for all $\varepsilon\in{\cal Q}_f$, we have $]a,b[\ \backslash{\cal I}_{\varepsilon}\neq\emptyset$. Therefore for all $\varepsilon\in{\cal Q}_f$ there exists $x\in]a,b[\backslash{\cal I}_{\varepsilon}$ such that $\Delta_{\varepsilon} f(x)\neq0$ (since $x\not\in {\cal I}_{\varepsilon}$), which gives that $f$ is a nowhere differentiable function on $]a,b[$ by Corollary \ref{Cor1}.

\rightline\Box

The converse uncertainty theorem can be formulated in terms of limitation as follow.

\begin{cor} (Converse Uncertainty Principle)
Let $f\in C^0([a,b])$. If it is impossible to have the following situations:

i) there exists $\varepsilon_0\in{\cal Q}_f$ such that for all $x\in]a,b[\ \backslash{\cal I}_{\varepsilon_0}$, $\Delta_{\varepsilon_0} f(x)$ and $\Delta_{\varepsilon_0} P(x)$ are well defined to an accuracy equal to \quad$\Delta_{\varepsilon_0} f(x)\ \Delta_{\varepsilon_0} P(x)= \varepsilon_0$,

ii) there exists $x_0\in\ ]a,b[\ \backslash \cup_{\varepsilon\in{\cal Q}_f}{\cal I}_{\varepsilon}$ such that for all $\varepsilon\in{\cal Q}_f$, $\Delta_\varepsilon f(x_0)$ and $\Delta_\varepsilon P(x_0)$ are well defined to an accuracy equal to \quad$\Delta_\varepsilon f(x_0)\ \Delta_\varepsilon P(x_0)= \varepsilon$,

iii) there exists $\varepsilon_0\in{\cal Q}_f$, there exists $x_0\in]a,b[\ \backslash{\cal I}_{\varepsilon_0}$ such that $\Delta_{\varepsilon_0} f(x_0)$ and $\Delta_{\varepsilon_0} P(x_0)$ are well defined to an accuracy equal to\quad $\Delta_{\varepsilon_0} f(x_0)\ \Delta_{\varepsilon_0} P(x_0)= \varepsilon_0$,

\ni then the function $f$ is nowhere differentiable on $]a,b[$.

\end{cor}

\subsection{General Interpretation of the Converse Uncertainty}

Inversely, the theorem \ref{IUP} asserts that if it is impossible to simultaneously determine the deviation $\Delta_\varepsilon f(t)$ and the normal average rate of change $\Delta_{\varepsilon} P(t)$ of a physical system to an accuracy equal to (\ref{IP}), and if the position function of the physical system is continuous, then it must be a nowhere differentiable function. The inverse of the uncertainty does not mean that the physical system doesn't have a position function if we are facing the impossibility in determining simultaneously the deviation of the position function and the rate of change of this deviation in a given continuous medium. It does rather have one, but it is unreachable, and more precisely it must be nowhere differentiable. Feynman and Hibbs \cite{FH} have proved that generic trajectories of quantum particles are continuous and nowhere differentiable, where their quadratic velocity exists and is given by the following limit
\begin{equation}
lim_{x\rightarrow x'} {(f(x)-f(x'))^2\over x-x'}.
\end{equation}
Moreover for the Brownian motion, Einstein \cite{AA} has proved that
\begin{equation}
f(x+h)-f(x)\approx h^{1\over2}\qquad \hbox{for} \qquad h> 0
\end{equation}
which reflects the non differentiability of the trajectory.
The set of continuous functions with non zero quadratic velocity is included in the set of continuous and nowhere differentiable functions, which makes this approach of the uncertainty principle and its converse using the nowhere differentiable functions valid for a quantum physical system that is subject to this indetermination, and makes the set of nowhere differentiable functions candidate for the quantum world. This approach via the nowhere differentiable functions provides another explanation of the need to change the use of the classical notion of single trajectory (for a given physical system) for the use of all possible trajectories (infinity of possible trajectories) to formulate the probabilistic amplitude in  Feynman's path integral \cite{FH2}.

\section{Prospect}

The equation (\ref{F5}) introduced in Remark \ref{L1bis} provides new information other than where lie the limits of uncertainty in measuring physical quantities. Indeed the equality (\ref{F5}) provides an equality between two different quantities. From one side (the left side of the equation (\ref{F5})) it is question of an unreachable position function $f(x)$ of one variable  that is characterized by the properties of a function everywhere continuous and nowhere differentiable, and from the other side (the right side of the equation (\ref{F5})) a linear combination and operations on a reachable position functions $f^+(x,\varepsilon)$ of two variables characterized by the properties of continuous and everywhere differentiable functions. This equality becomes wrong if we fix one variable in the mean function $f^+(x,\varepsilon)$, and it remains always valid by considering a function of two variables $f^+(x,\varepsilon)$ defined on  $]a,b[\times {\cal Q}_f$. The equation (\ref{F5}) says that if we use a surface that contains all possible approximations of the real position function $f(x)$, the unreachable can be reached and an exact result concerning the nowhere differentiable position function can be obtained via operation on that surface.

This means that the consideration of an infinite continuous family of approximations (represented in the surface $z=f^+(x,\varepsilon)$) is needed to obtain an accurate description of the unreachable nowhere differentiable function. The simultaneous determination of the deviation of the position function $\Delta_{\varepsilon} f(x)$ and the rate of change of this deviation $\Delta_{\varepsilon} P(x)$ to an accuracy equal to (\ref{IP}) is only subject to the regularity of the position function $f$.

We can always use the quantum mechanics framework and formalism in hunting the density of probability of the position function $f(x)$, however a new framework based on the property conveyed by the equation (\ref{F5}) that contains the uncertainty relation might provide an alternative formalism that could enrich the quantum mechanics formalism using properties of nowhere differentiable functions. Some attempts in this direction have been elaborated and have led to obtain a new kind of differentiable manifold called fractal manifold (\cite{BF, BF3}) that presents variable geometry and variable topology \cite{P}, and this framework has also recently led to found an inner connection between gravity and electromagnetism \cite{BF4}.

\bibliographystyle{spmpsci}
\bibliography{allbiblio}

\end{document}